\pdfoutput=1
\documentclass[a4paper,american,floatfix,pdftex,superscriptaddress,twoside,%
aps,pra,
reprint,
]{revtex4-2}%
\usepackage{amsfonts,amsmath,amssymb}
\usepackage[T1]{fontenc}
\usepackage{graphicx}%
\usepackage[utf8]{inputenc}
\usepackage{mathtools}
\usepackage{hyperref, hypernat}
\usepackage[todos]{CMImacros} 
\usepackage[displaymath,textmath,graphics]{preview}

\graphicspath{{./figures/}} 

\newcommand*{\dds}{\ensuremath{\text{D}_2\text{S}}\xspace}%
\newcommand*{\deuterosulfide}{\ensuremath{\text{dihydrogen sulfide}}\xspace}%
\newcommand*{\imi}{\ensuremath{\text{C}_3\text{H}_4\text{N}_2}\xspace}%
\newcommand*{\imidazole}{\ensuremath{\text{2H-imidazole}}\xspace}%

\newcommand{\cfeldesy}{\affiliation{Center for Free-Electron Laser Science, Deutsches
      Elektronen-Synchrotron DESY, Notkestraße 85, 22607 Hamburg, Germany}}%
\newcommand{\uhhcui}{\affiliation{Center for Ultrafast Imaging, Universität Hamburg, Luruper
      Chaussee 149, 22761 Hamburg, Germany}}%
\newcommand{\uhhphys}{\affiliation{Department of Physics, Universität Hamburg, Luruper Chaussee 149,
      22761 Hamburg, Germany}}%
	\newcommand{\jkemail}{\email[Email:~]{jochen.kuepper@cfel.de}}%
\newcommand{\ezemail}{\email{emil.zak@cfel.de}}%
\newcommand{\cmiweb}{\homepage[\\website:~]{https://www.controlled-molecule-imaging.org}}%

\begin{document}
\title{Controlling rotation in the molecular-frame with an optical centrifuge}%
\author{Emil J.\ Zak}\ezemail\cfeldesy%
\author{Andrey Yachmenev}\cfeldesy\uhhcui%
\author{Jochen Küpper}\jkemail\cmiweb\cfeldesy\uhhcui\uhhphys%
\date{\today}%
\begin{abstract}\noindent%
   We computationally demonstrate a new method for coherently controlling the rotation-axis
   direction in asymmetric top molecules with an optical centrifuge. Appropriately chosen
   electric-field strengths and the centrifuge's acceleration rate allow to generate a nearly
   arbitrary rotational wavepacket. For \deuterosulfide (\dds) and \imidazole (\imi) we created
   wavepackets at large values of the rotational quantum number $J$ with the desired projections of
   the total angular momentum onto two of the molecules' principal axes of inertia. One application
   of the new method is three-dimensional alignment with a molecular axis aligned along the laser's
   wave vector, which is important for the three-dimensional imaging of molecules yet not accessible
   in standard approaches. The simultaneous orientation of the angular momentum in the laboratory
   frame and in the molecular frame could also be used in robust control of scattering experiments.
\end{abstract}
\maketitle%

\section{Introduction}
Preparing well-defined molecular samples, maximally fixed in three-dimensional space, has been a
long-term goal in physics, chemistry, and related fields such as quantum
technology~\cite{Brooks:Science193:11, Friedrich:PRL74:4623, Stapelfeldt:RMP75:543,
   Koch:RMP91:035005}. Significant progress toward this goal was made in the gas phase using
internally cold molecular samples and tailored external fields~\cite{Krems:ColdMolecules,
   Holmegaard:PRL102:023001, Ghafur:NatPhys5:289, Chang:IRPC34:557}. Nowadays, it is possible to
strongly align and orient molecules in the laboratory frame in one~\cite{RoscaPruna:PRL87:153902,
   Chatterley:JCP148:221105, Karamatskos:NatComm10:3364, Fleischer:PRL107:163603},
two~\cite{Smeenk:PRL112:253001, Smeenk:JPB46:201001, Korobenko:PRL116:183001}, and
three~\cite{Larsen:PRL85:2470, Tanji:PRA72:063401, Rouzee:PRA77:043412, Nevo:PCCP11:9912,
   Lin:NatComm9:5134} spatial directions. It is also possible to create unidirectionally rotating
molecules with oriented-in-space angular momentum and a narrow rotational energy
spread~\cite{Zhdanovich:PRL107:243004, Milner:ACP159:395, Milner:PRA93:053408, Owens:PRL121:193201}.
Such alignment and orientation of molecules is critical for imaging molecular structure and
ultrafast dynamics in the molecule-fixed frame~\cite{Filsinger:PCCP13:2076, Itatani:Nature432:867,
   Meckel:Science320:1478, Holmegaard:NatPhys6:428, Hensley:PRL109:133202, Barty:ARPC64:415,
   Kuepper:PRL112:083002, Yang:Science361:64} as well as in the stereodynamical control of chemical
reactions~\cite{Larsen:PRL83:1123, Miranda:NatPhys7:502, Liu:ARPC52:139, Shagam:NatChem7:921}.

A highly efficient technique used to generate and control the molecule's angular momentum is the
optical centrifuge~\cite{Karczmarek:PRL82:3420, Villeneuve:PRL85:542}, which is a strong
non-resonant linearly polarized laser pulse that performs accelerated rotation of its polarization
about the direction of propagation. It can excite molecules into rotational states with extremely
large angular momentum, creating an ensemble of superrotors~\cite{Yuan:PNAS108:6872,
   Korobenko:PRL112:113004}. Molecules in superrotor states are aligned in the polarization plane of
the centrifuge~\cite{Milner:PRA93:053408} and resist collisional decoherence for
microseconds~\cite{Yuan:PNAS108:6872, Khodorkovsky:NatComm6:7791, Milner:PRL113:043005,
   Milner:PRX5:031041}.

The optical centrifuge can also be utilized as a versatile tool to fine tune the rotational dynamics
of molecules, including the coherent control of the rotation axis~\cite{Owens:JPCL9:4206}, the
enantiomer-specific excitations of chiral molecules~\cite{Tutunnikov:JPCL9:1105,
   Milner:PRL122:223201, Yachmenev:PRL123:243202}, and even the creation of chiral samples from
achiral molecules~\cite{Owens:PRL121:193201}.

Here, we further explore the possibility of tailoring the optical field to steer the rotational
dynamics of asymmetric top molecules. We computationally demonstrate a new type of
rotational coherent control by exciting the rotation of an asymmetric top molecule about two
different axes of inertia with simultaneously fully controlled orientation of the angular momentum
in both, the laboratory-fixed and the molecule-fixed, frames.

We apply our method to asymmetric top molecule \dds, which exhibits the effects of rotational
energy level clustering and dynamical chirality at high rotational excitations
\cite{Owens:JPCL9:4206, Owens:PRL121:193201,Bunker:JMolSpec228:640}. To
populate the rotational cluster states, \dds has to be excited along a specific pathway of
rotational states. Previously, we explored the method of pulse shaping, \ie, by repeatedly
turning on and off the field, to make the asymmetric top molecule H$_2$S rotate about either of
its two stable inertial axes. We demonstrated that only the $a$-axis rotational excitation leads
to the population of rotational cluster states in H$_2$S. Here, we propose a more robust
approach to create \emph{arbitrary} coherences between the two stable molecular rotations, which
relies only on the careful selection of the centrifuge angular acceleration rate. The heavier
\dds isotopologue is chosen for the present study due to the lower laser intensities required for
efficient rotational excitation, thus reducing potential ionization. Along with robust
quantum-mechanical calculations, we derived a simple analytical metric that allows to predict the
orientation of the angular momentum in the molecule-fixed frame for arbitrary molecules given the
parameters of the optical centrifuge.

Additionally, by adjusting the turn-off time we demonstrate the two types of three-dimensional
(3D) alignment of asymmetric top molecules, with either of the two stable rotation axes pointing
along the field's wave-vector, so called
$k$-alignment~\cite{Smeenk:PRL112:253001, Smeenk:JPB46:201001,
	Pickering:PRA99:043403}.
\begin{figure}
   \includegraphics[width=\linewidth]{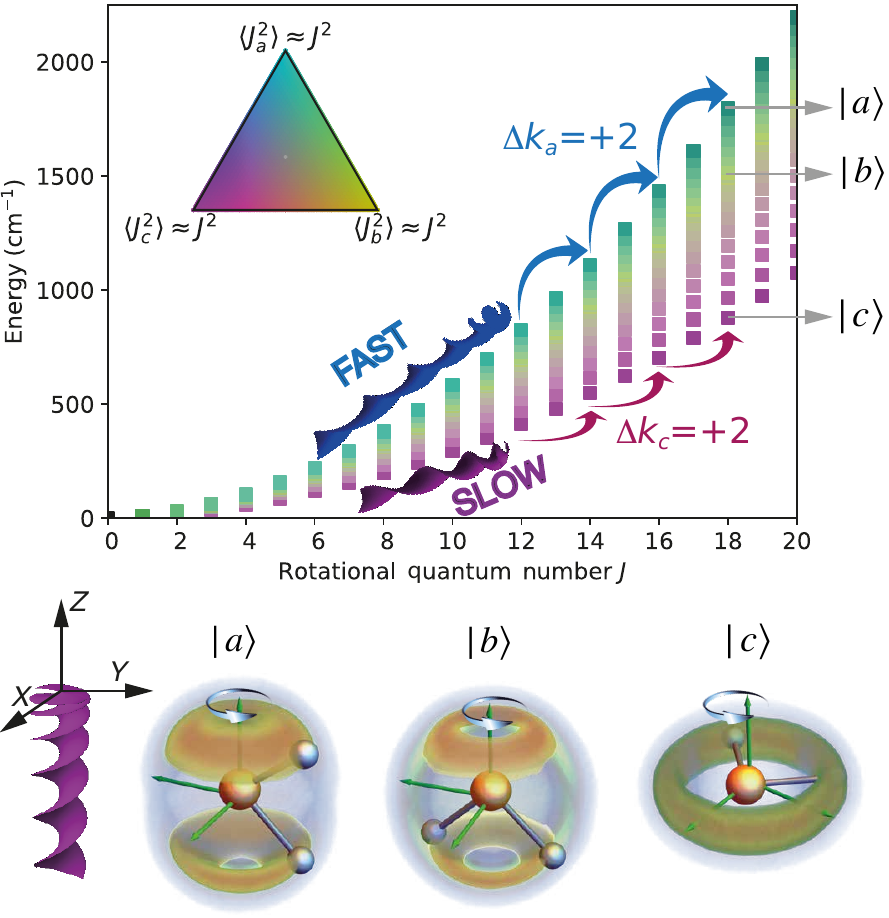}%
   \caption{Top panel: rotational-energy-levels of the near-oblate asymmetric top molecule \dds as a
      function of the rotational quantum number $J$. Each level is color-coded depending on its
      expectation values $\expectation{\hat{J}_a^2}$, $\expectation{\hat{J}_b^2}$, and
      $\expectation{\hat{J}_c^2}$ according to the color map shown in the inset. Arrows
      schematically mark the two competing excitation pathways ``fast'' and ``slow''. Three
      principal rotation states for $J=18,M=18$ are denoted as $\ket{a}, \ket{b}$ and $\ket{c}$; see
      text for details. Bottom panel: 3D probability-density distributions of the deuterium atoms
      for three selected principal rotation states. In the laboratory frame, the centrifuge pulses
      propagate along the laboratory $Z$ axis, trap two of the molecule-fixed axes in the $XY$ plane
      and unidirectionally spin the molecules about the third axis, which is aligned along $Z$.}
   \label{fig:energy-levels}
\end{figure}

\section{Principal-rotation states}
\autoref{fig:energy-levels} displays the rotational energy level structure of the \dds molecule. For
any value of the rotational quantum number $J$ there is a multiplet of $2J+1$ levels, which are
colored according to the average values of the angular momentum projection operators onto the
principal axes of inertia $\expectation{\hat{J}_a^2}$, $\expectation{\hat{J}_b^2}$, and
$\expectation{\hat{J}_c^2}$. The \hyperref[fig:energy-levels]{bottom panel} in \autoref{fig:energy-levels} 
shows the calculated 3D
probability density for deuterium atoms for the highest-, middle-, and lowest-energy levels at
$J=18, M=18$. Here, $M$ is the quantum number for the $Z$-component of the angular momentum operator
in the laboratory-fixed frame. It is evident that the highest-energy levels within each $J$
multiplet correspond to $\expectation{\hat{J}_a^2}\approx J^2$ (cyan color), \ie, in these states
the molecule rotates about the $a$-axis and $k_a=J$ becomes a near-good quantum number. The
lowest-energy levels correspond to rotation about the $c$-axis with $k_c=J$ (purple color), while those with energies in the middle are
mixtures of
rotations about different axes with some of them exhibiting classically unstable {$b$-axis rotation}
(yellow color).

We refer to \emph{principal-rotation states} when the rotational angular momentum is nearly aligned
along one of the principal axis of inertia, \ie, $k_a=J$, $k_c=J$, or $k_b=J$. Controlling
populations of the principal rotation states, allows to create arbitrary three-dimensional
orientation of the total angular momentum in the molecule-fixed frame. We investigated the
orientation of the angular momentum in the $ac$ plane by controlling the populations of the lowest-
and highest-energy-state components in the rotational wavepacket.

Principal rotation states can be populated through the interaction with the optical centrifuge
field, represented by
\begin{align}
\mathbf{E}(t) = E_0 f(t) \cos(\omega t)\left[\mathbf{e}_X\cos(\beta t^2)+\mathbf{e}_Y\sin(\beta t^2) \right],
\end{align}
with the peak amplitude $E_0$, the pulse envelope $f(t)$, the acceleration $\beta$ of
angular rotation of the polarization, and the far off-resonant ($\lambda=800$~nm) carrier
frequency $\omega$ of the linearly polarized pulse. A molecule placed in the optical centrifuge field experiences a series of
Raman transitions with
$\Delta{J}=2$ and $\Delta{m}=\pm2$, depending on the sign of $\beta$, which defines the direction of
the centrifuge rotation.

Initially in the rotational ground state~\cite{Chang:IRPC34:557}, the molecule can undergo two main
excitation pathways: along the lowest-energy or along the highest-energy rotational states of the
$J$ multiplets. Excitation rates for these two pathways are governed by the two specific
polarizability-interaction terms in the molecule-field interaction potential:
\begin{equation}
\begin{split}
V(t) =& \frac{1}{4}E_0^2 f(t)^2 \cos^2(\omega t)\left[ \right.\\
		&-\Delta \alpha_1 \left( D_{2,0}^{(2)*}e^{-2i\beta t^2} + D_{-2,0}^{(2)*}e^{2i\beta t^2} -\sqrt{\frac{2}{3}}D_{0,0}^{(2)*}\right) \\
		&\left. -\Delta\alpha_2 \left( S_{2,2}^{(2)*}e^{-2i\beta t^2} + S_{-2,2}^{(2)*}e^{2i\beta t^2} -\sqrt{\frac{2}{3}}S_{0,2}^{(2)*} \right)
		\right]
\end{split}
\label{eq:potential}
\end{equation}
Here, $D_{M,k}^{(J)*}$ denotes the complex-conjugated Wigner $D$-matrix and
$S_{M,k}^{(2)*}=D_{M,k}^{(2)*}+D_{M,-k}^{(2)*}$.  The expressions for $\Delta\alpha_1$ and
$\Delta\alpha_2$ depend on the choice of the quantization axis in the molecule-fixed frame.

A molecule is called (near) prolate if the quantization axis is along the $a$ axis and (near) oblate
if it is along the $c$ axis. The first term in \eqref{eq:potential} describes the interaction with
the polarizability anisotropy,
$\Delta\alpha_1=\frac{1}{\sqrt{6}}(2\alpha_{aa}-\alpha_{bb}-\alpha_{cc})$ for (near-)prolate and
$\Delta\alpha_1=\frac{1}{\sqrt{6}}(2\alpha_{cc}-\alpha_{aa}-\alpha_{bb})$ for (near-)oblate top.
This term yields $\Delta k_a=0$ transitions along the lowest-energy pathway of rotational states for
(near-)prolate top molecules and $\Delta k_c=0$ transitions along the highest-energy pathway for
(near-)oblate top molecules. The second term in \eqref{eq:potential}, with
$\Delta\alpha_2=\frac{1}{2}(\alpha_{bb}-\alpha_{cc})$ for (near-)prolate and
$\Delta\alpha_2=\frac{1}{2}(\alpha_{aa}-\alpha_{bb})$ for (near-)oblate top, gives $|\Delta k_a|=2$
and $|\Delta k_c|=2$ transitions corresponding to the highest-energy and lowest-energy pathways for
\mbox{(near-)}prolate and (near-)oblate top molecules, respectively.
\begin{figure}
   \centering%
   \includegraphics[width=0.7\linewidth]{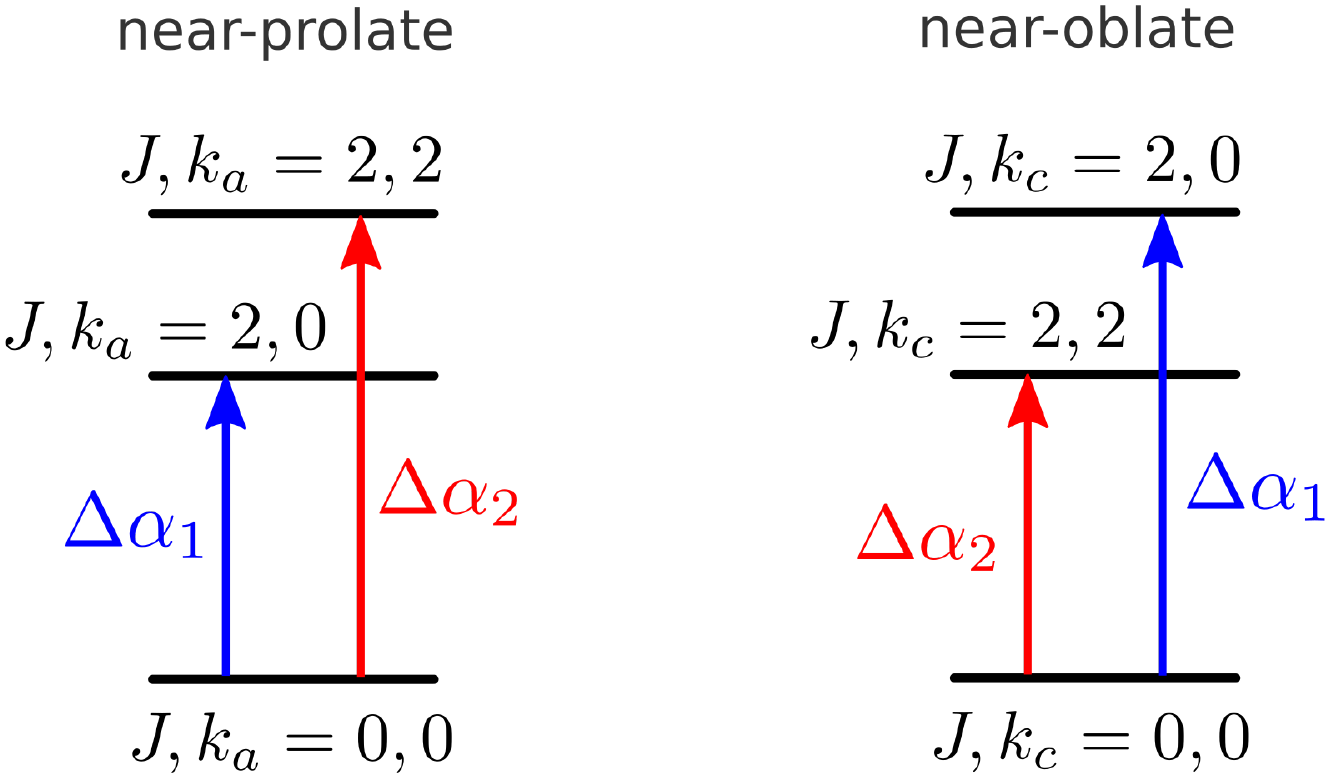}%
   \caption{Scheme of the centrifuge driven transitions from $J=0$ to $J=2$ states in near-prolate
   	and near-oblate top molecules. The symbols $\Delta\alpha_1$ and $\Delta\alpha_2$  represent the transition moments
   	defined in \eqref{eq:potential}.}
   \label{fig:transitions}
\end{figure}

The ability to control the excitation pathway and ultimately the orientation of the angular momentum
in the molecule-fixed $ac$ plane depends on the relationship between the molecular polarizability
anisotropies $\Delta\alpha_1$ and $\Delta\alpha_2$, see \autoref{fig:transitions}. For a great
majority of molecules it applies that $\abs{\Delta\alpha_1}>\abs{\Delta\alpha_2}$, and we assume
this in the discussion below. Hence, for a near-prolate top molecule, the transition moment into the
lowest-energy state is greater than that into the highest-energy state and \emph{vice versa} for a
near-oblate top molecule.

A molecule in the optical centrifuge climbs the rotational energy level ladder via a series of
consecutive $\Delta J=2$ excitations between neighboring rotational states with $J=0,2,4,\ldots$
Starting from $J=0$, the resonance of the centrifuge's frequency $\omega(t) = \beta t$ with the lowest-energy state in the
$J=2$ multiplet occurs earlier in time than the resonance with the highest-energy state in the same
multiplet. Thus, in order to steer the excitation along the highest-energy pathway in the
near-prolate case one would suppress the stronger lower-frequency transitions by applying shaped
pulses, \ie, by repeatedly decreasing and increasing the field intensity $f(t)$ at the crossing
times with unwanted and desired transitions, respectively~\cite{Owens:JPCL9:4206}. In the
near-oblate case the highest-energy excitation pathway is favorable. However, it is possible to
guide it along the lowest-energy path by only adjusting the centrifuge peak field strength $E_0$ and
acceleration rate $\beta$. In light of recent experiments~\cite{MacPhailBartley:RSI91:045122} such
an approach seems more feasible than the pulse-intensity shaping strategy.

We investigate the possibility of controlling the rotational wavepacket composition in near-oblate
top molecules by selectively populating the \emph{principal rotation states} using the optical
centrifuge with appropriately chosen intensity and acceleration rate $\beta$. For small $\beta$
values and high intensities the centrifuge's rotating field will first slowly cross through
resonance with the ground- to lowest-excited rotational-energy level transition. If the
corresponding transition moment $\ordsim\abs{\Delta\alpha_2}^2$ is not entirely negligible, it will
predominantly populate the $c$-axis principal rotation states. On the other hand, for large $\beta$
values and low intensities the centrifuge will chirp through resonance with low-energy $c$-axis
principal rotation states fast enough to not populate them significantly. As a result the stronger
transition to the $a$-axis principal rotation states will dominate, see \autoref{fig:transitions}
and \appautoref{app-A} for a more detailed discussion. By choosing intermediate values for
$\beta$, an arbitrary coherent rotational wavepacket over $a$- and $c$-axis principal rotation
states can be tailored.

\section{Computational details}
We computationally demonstrate the proposed technique for the near-oblate asymmetric top
molecules \dds and \imidazole (\imi). The latter has been chosen to examine the effectiveness of
our technique for larger molecules, \ie, with rotational constants smaller than in \dds and
larger electronic polarizabilities.Our calculations employed a
highly-accurate
variational approach. The rotational-dynamics calculations of \dds and \imidazole were performed in two steps. In the
first step, the molecular field-free energies and their transition moments were obtained. For D$_2$S
we utilized the full-dimensional variational procedure TROVE~\cite{Yurchenko:JMS245:126,
	Yachmenev:JCP143:014105, Yurchenko:JCTC13:4368} together with a highly-accurate spectroscopically
adjusted potential energy surface~\cite{Azzam:MNRAS460:4063} and high-level \emph{ab initio}
polarizability surface~\cite{Owens:JPCL9:4206} of the H$_2$S molecule within the Born-Oppenheimer
approximation. The field-free basis for \imidazole molecule was produced using the rigid-rotor
approximation with the rotational constants calculated from the equilibrium geometry, obtained using
density functional theory (DFT) with the B3LYP functional and the def2-QZVPP basis
set~\cite{Weigend:JCP119:12753,Weigend:PCCP7:3297}. In the second step, the time-dependent solutions
for the full molecule-field interaction Hamiltonian were obtained using the computational approach
Richmol~\cite{Owens:JCP148:124102,Yachmenev:JCP151:244118}. The wavefunctions were time-propagated
using the split-operator method with a timestep of 10~fs. The time-evolution operator was evaluated
using an iterative approximation based on the Krylov subspace methods.
\begin{figure}
   \includegraphics[width=0.9\linewidth]{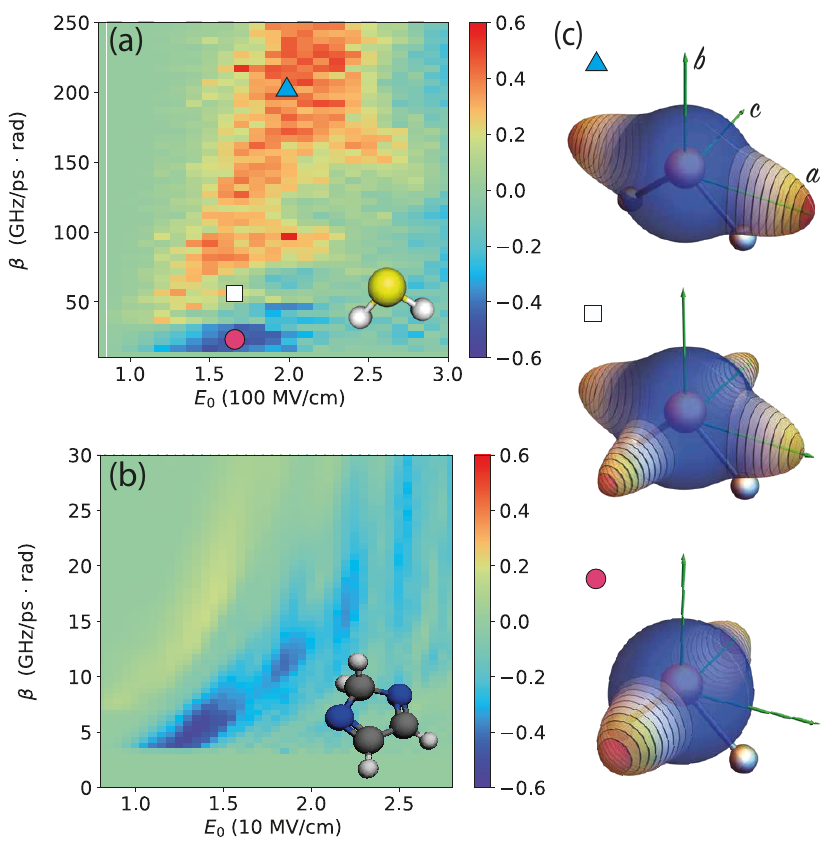}%
   \caption{Differences in cumulative populations of the \ket{c} and \ket{a} principal rotation
      states as functions of the centrifuge's peak field strength $E_0$ and acceleration $\beta$,
      plotted in (a) for \dds and in (b) for \imidazole; see also~\autoref{fig:results}. Panel (c) displays 3D
      probability distributions for the molecular-frame rotation axes of \dds, showing $a$, $a/c$,
      and $c$ rotation generated with three different sets of centrifuge parameters, marked in (a)
      by a triangle, a square, and a circle, respectively.}
   \label{fig:populations}
\end{figure}

\section{results and discussion}
The calculated cumulative population-inversion between \ket{a} and \ket{c} principal rotation states
for \dds and \imidazole are shown in \autoref{fig:populations}; see \autoref{fig:results} in  \appautoref{app-B}
 for the individual populations.
The cumulative population-inversion was calculated as the difference between the state
populations, in field-free conditions after the 150~ps centrifuge pulse was turned off, summed
along the \ket{a} and the \ket{c} excitation paths for $J\geq10$ for \dds and for $J\geq20$ for
\imidazole.

As long as the strong ro-vibrational coupling effects do not break the $\ket{a}$ or $\ket{c}$
excitation chain, the final distribution of populations across the states with different $J$ can
be controlled by the centrifuge turn-off time. In principle, super-rotor states can be populated.
For rigid-rotor \imidazole the rotational excitation proceeds unhindered up to high angular
momentum states. For \dds, the $\ket{a}$ rotational-excitation chain breaks around $J=20$ due to
the centrifugal distortion effects characteristic for the molecules with rotational energy level
clustering \cite{Kozin:JCP104:4105}. Thus, as a measure of the rotational excitation yield, we
use the cumulative populations of states with $J\geq20$ for \imidazole and $J\geq10$ for \dds.

The populations plotted in \autoref{fig:populations} are functions of the centrifuge peak field $E_0$ and the rate of
acceleration $\beta$. In both molecules, small acceleration rates populate mainly the \ket{c}
states. With increasing acceleration rate the optical centrifuge populates more preferably the
\ket{a} states. The relation between the wavepacket composition and the probability of the
rotation-axis orientation in the molecular frame is displayed in \autoref[c]{fig:populations} for
three selected sets of the centrifuge parameters. A closer inspection of
\autoref[a]{fig:populations} reveals that at small acceleration rates \betaapproxrate{50} solely
changing the field from weak to strong switches the created wavepacket from \ket{a}-dominated to
\ket{c}-dominated. After initial $\ket{a}$/$\ket{c}$ bifurcation for $J=0\rightarrow2$ the
rotational excitation proceeds nearly loss-free in \dds and, due to its higher density of states,
with some losses in \imidazole.

The preference of an asymmetric top molecule to rotate about the $a$- or $c$- principal axis of
inertia is determined by its polarizability and rotational constants, which define the transition
moments between rotational states, and the properties of the centrifuge field. All above quantities
can be represented in a vector
\begin{equation}
   S_i \coloneqq R_i (Q_{ji}+Q_{ki})  \qquad i,j,k=a,b,c
   \label{eq:rotability}
\end{equation}
where $i\neq{j}\neq{k}\neq{i}$ label the principal axes of inertia. $R_i$ describes the molecule's
ability to rotate about an axis $i=a,b,c$ and is defined as
$R_i=C_i(|\alpha_{jj}-\alpha_{kk}|)/(C_j(|\alpha_{kk}-\alpha_{ii}|)+C_k(|\alpha_{jj}-\alpha_{ii}|))$,
where $\alpha_{ii}$ are the diagonal components of the electronic polarizability and $C_i=A,B,C$ for
$i=a,b,c$. $Q_{ij}$ describes the quantum-mechanical population
\emph{transferability} defined as $Q_{ij}=P_i/P_j$, with the Landau-Zener populations
$P_i=1-\exp\left(-\pi\Omega_i^2/(4\beta)\right)$~\cite{Vitanov:PRA53:4288} and the Rabi frequency
$\Omega_i=\mu_{i}E_0^2$, which depends on the transition moment from the rotational ground state to
one of the \ket{i} ($i=a,b,c$) principal rotation states and $E_0$ is the electric field strength.

The \emph{rotability} $\vec{S}$ \eqref{eq:rotability} quantifies the molecule's preference to rotate
about the different inertial axes. It is composed of the two quantities: 1) The \emph{trapability}
$R_i$ measures the molecule's capability to form certain pendular bound-states in a centrifuge
field, \ie, it provides a measure of how confined the molecule is in the pendular potential well
along each of the principal axes. 2) The quantum \emph{transferability} $Q_{ij}$ accounts for the
transition moment differences between the respective $\ket{a}$, $\ket{b}$, $\ket{c}$ principal
rotational excitation branches. It quantifies the preference of the system to choose one rotational
excitation path over the other, as a function of the optical centrifuge parameters $\beta$ and
$E_0$, see  \appautoref{app-C} for details.

\begin{figure}
   \includegraphics[width=\linewidth]{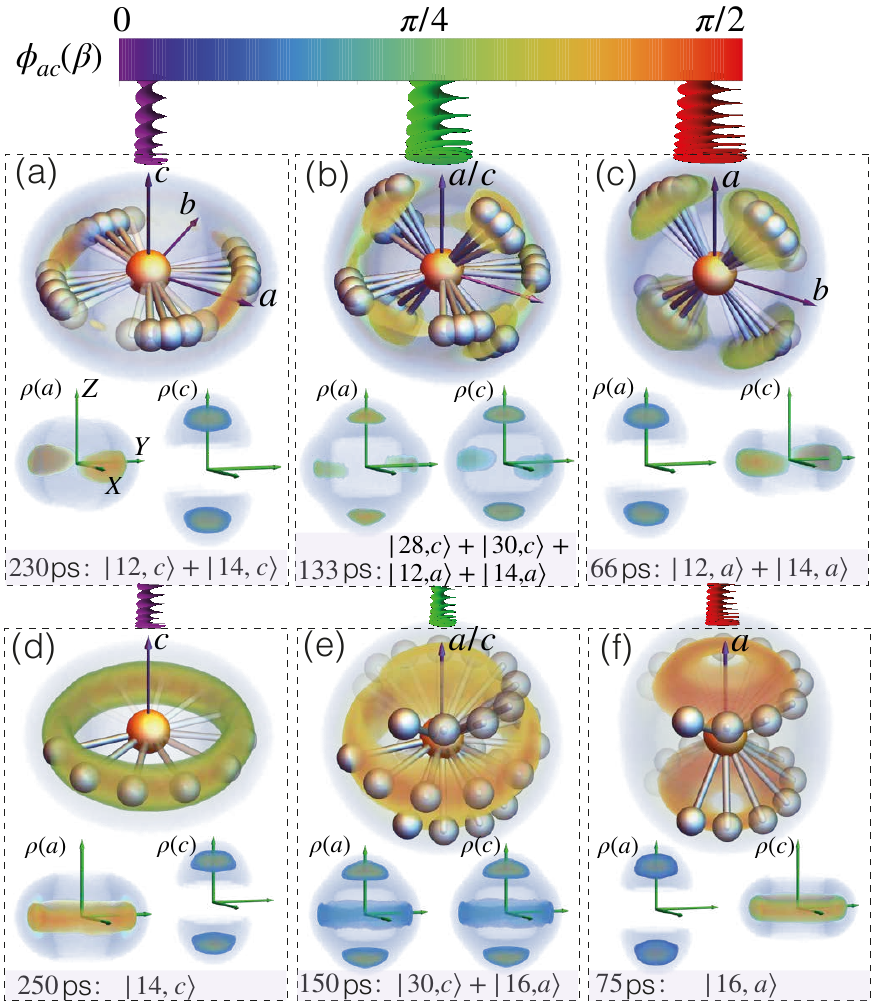}%
   \caption{3D probability density distributions of the deuterium nuclei and the $a$ and $c$ axis,
      $\rho(a)$ and $\rho(c)$, of \dds. The deuterium-nuclei density plots shown in the upper part
      of each panel provide a complementary view to the inertial-axis density plots displayed in the
      panels' bottom part. These distributions are depicted for wavepackets generated with an
      optical centrifuge at three different acceleration rates \betarate{30,115,230} and field
      strength $E_0=2\cdot10^{8}$~V/cm. The different centrifuge-field turn-off times and the
      resulting wavepacket compositions $\ket{J,a/c}$ are specified below each plot. The colorbar
      represents values of the angle $\phi_{ac}=\arctan\left(S_a/S_c\right)$, which correlates with
      the centrifuge's acceleration rate; see text for details. (a--c) wavepackets showing different
      types of 3D alignment; (d--f) wavepackets without 3D alignment. The $XYZ$ coordinate system in
      (a) indicates the laboratory frame. Relevant principal inertial axes are marked in purple.
      Blurred atoms reflect the probability density of finding the deuterium and sulfur atoms in 3D
      space.}
   \label{fig:PDF}
\end{figure}
\autoref{fig:PDF} shows the 3D rotational probability density distributions for the deuterium nuclei
in \dds for wavepackets generated by the optical centrifuge at electric field strength
$E_0=2\cdot10^{8}$~V/cm and three different acceleration rates: \betarate{30,115,230}. The
acceleration rates are conveniently encoded in colorbar representing the angle
$\phi_{ac}=\arctan\left(S_a/S_c\right)$ which relates the $a$ and $c$ elements of the rotability
vector $\vec{S}$ given in \eqref{eq:rotability}. Wavepackets simulated at $\phi_{ac} \approx 0$,
\autoref[a,d]{fig:PDF}, $\phi_{ac} \approx \pi/4$, \autoref[b,e]{fig:PDF}, and
$\phi_{ac} \approx \pi/2$, \autoref[c,f]{fig:PDF}, display strong characteristics of $\ket{c}$,
$\ket{c}+\ket{a}$ and $\ket{a}$ states, respectively; \cf \autoref{fig:energy-levels}.

At low acceleration rates $\beta$ the rotability vector is dominated by the $S_c$ component
($\phi_{ac} \approx 0$) and the rotational wavepacket mainly consists of $|c\rangle$ states; the
deuterium atoms' 3D probability forms a ring shown in \autoref[d]{fig:PDF}. With increasing $\beta$
the ratio $S_a/S_c$ also increases and the wavepacket composition smoothly converts to the
$|a\rangle$-state dominant ($\phi_{ac} \approx \pi/2$) in \autoref[f]{fig:PDF}. To summarize, for a
given set of rotational constants, electronic polarizabilities, and corresponding transition dipole
moments, the rotability vector depends solely on the acceleration rate of the centrifuge $\beta$ and
its electric field strength $E_0$. Thus, this rotability vector is a very useful quantity for
estimating the centrifuge parameters needed to reach a desired $\ket{a}/\ket{c}$-composition of the
wavepacket without the need for costly quantum-mechanical computations.

In addition, depending on centrifuge's turn-off time shown below each plot in \autoref{fig:PDF}, the
end-product wavepacket can become dominated by principal rotation states (or mixtures of those)
either with a single $J$ value or with $J$,$J+2$ coherences. In the latter case, the respective
probability densities evolve in time (see \appautoref{app-D} and supplementary materials).
Snapshots of such wavepackets are displayed in \autoref[a-c]{fig:PDF}. A high degree of 3D alignment
is visible in these wavepackets. In supplementary materials we also show simulated velocity-map
images which record $D^+$ ions after Coulomb exploding the molecules - an experiment which can
detect and characterize different principlal rotation states.

The new protocol for controlling populations of the $\ket{a}$ and $\ket{c}$ rotational states opens
an avenue~\cite{Smeenk:JPB46:201001, Korobenko:PRL116:183001, Korobenko:PCCP17:951} to 3D
aligning~\cite{Larsen:PRL85:2470, Nevo:PCCP11:9912, Kierspel:JPB48:204002} molecules with either
their largest or smallest polarizability axis pointing along the wave-vector of the alignment laser
(centrifuge). Such $k$-alignment~\cite{Smeenk:JPB46:201001, Pickering:PRA99:043403} is desired in
many ultrafast imaging experiments~\cite{Holmegaard:NatPhys6:428, Smeenk:PRL112:253001,
   PopovaGorelova:PRA94:013412, Smirnova:Nature460:972, Trabattoni:NatComm11:2546}) yet so far has
not been realized for asymmetric top molecules. Here, we show that the centrifuge-field turn-off
time can be used to steer the 3D $k$-alignment of molecules. For example, the wavepacket shown in
\autoref[a]{fig:PDF} is dominated by a uniform mixture of \dds $\ket{J=12,c}$ and $\ket{J=14,c}$
states, it exhibits classical-like rotation of 3D-localized nuclear probability
density~\cite{Lapert:PRA83:013403}. The molecular $ab$ plane is confined in the $XY$ rotation plane
of the optical centrifuge and the $c$-axis aligned along the wave-vector $Z$ of the pulse. Ramping
up the centrifuge's acceleration rate to a high value of \betarate{230} ($\phi_{ac}\approx\pi/2$)
yields, after 66~ps, approximately a $\ket{J=12,a}+\ket{J=14,a}$ rotational wavepacket, depicted in
\autoref[c]{fig:PDF}, where this time the $a$ axis points along the light's wave-vector.
Interestingly, for an intermediate acceleration rate \betarate{115} ($\phi_{ac}\approx\pi/4$), a
superposition of 3D aligned states in which the $a$- and $c$- axis is simultaneously pointing along
the laboratory $Z$-axis is shown in the middle plot in \autoref[b]{fig:PDF}.

\section{summary}
In summary, we demonstrated that appropriate values of the optical centrifuge's acceleration rate
and intensity gauge the rotational wavepacket composition. Essentially arbitrary coherence between
the \ket{a} and \ket{c} principal rotation quantum states can be achieved. Through elementary use of
an optical centrifuge one can prepare molecular ensembles in the gas phase in which both the
laboratory-fixed angular momentum as well as the molecule-fixed angular momentum are robustly
controlled. By appropriately choosing the turn-off time of the centrifuge field one can also create
wavepackets that exhibit classical-like rotation and exhibit high degrees of 3D alignment with the
$a$ or $c$ axis pointing along the wave-vector of the driving field.

Natural applications of such tailored wavepackets, which are typically
long-lived~\cite{Yuan:PNAS108:6872}, are stereodynamics studies~\cite{Murray:JCP148:084310}, \eg, in
crossed-molecular-beams or surface-scattering experiments investigating collisional properties,
reactive scattering, or stereodynamical control of chemical reactions.

The presented method for coherent control of the rotation axis is expected to work best for light
asymmetric near-oblate-top molecules with large electronic polarizability anisotropies
$\Delta\alpha_1=2\alpha_{zz}-\alpha_{xx}-\alpha_{yy}$ and
$\Delta\alpha_2=\alpha_{xx}-\alpha_{yy}$. In the case of symmetric-top molecules with
$\Delta\alpha_2=0$, it is only possible to populate the lowest-energy rotational excitation path
for prolate tops and the highest-energy path for oblate tops. In light molecules, the lower
density of rotational states significantly aids the control of the excitation pathway and at the
same time is very well suited for the production of very cold molecular beams using the
deflection techniques \cite{Chang:IRPC34:557}. On the other hand, lighter molecules
generally require stronger fields to efficiently excite the Raman transitions, which can lead to
ionization depletion.

The proposed method has a tolerance for the fluctuation in the
laser intensity of about 20--25~\%. We point out that our simulations utilized experimental
parameters that are within the capabilities of typical present day laser and molecular beam
technology.

\section*{Acknowledgements}
We thank Stefanie Kerbstadt for fruitful discussions. This work has been supported by the Deutsche
Forschungsgemeinschaft (DFG) through the priority program ``Quantum Dynamics in Tailored Intense
Fields'' (QUTIF, SPP~1840, YA~610/1) and the Cluster of Excellence ``Advanced Imaging of Matter''
(AIM, EXC~2056, ID~390715994). We acknowledge support by Deutsches Elektronen-Synchrotron DESY, a
member of the Helmholtz Association (HGF), and the use of the Maxwell computational resources
operated at Deutsches Elektronen-Synchrotron DESY.

\appendix
\section{Simple model of the rotational-state-population branching}
\label{app-A}
\begin{figure}[b]
   \includegraphics[width=0.7\linewidth]{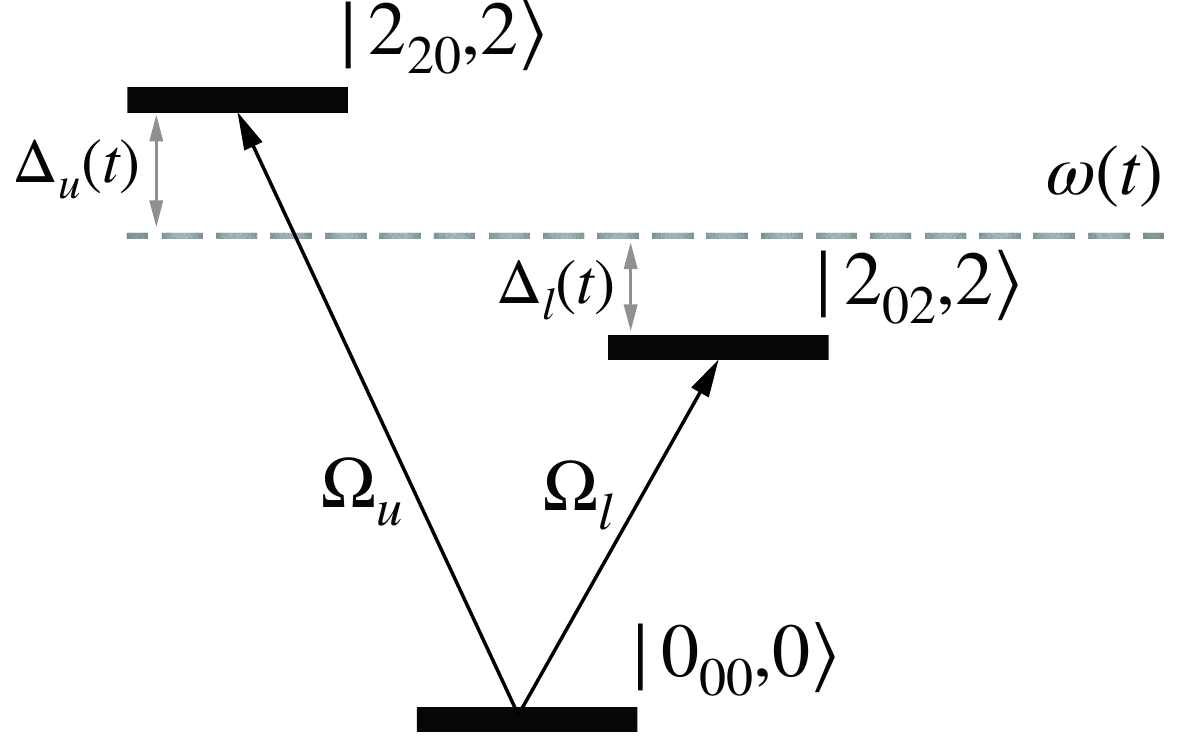}%
   \caption{A model 3-level system representing the lowest-energy rotational states in a near-oblate
      asymmetric top molecule. The levels are denoted as
      $\ket{J_{k_ak_c}, m} = \ket{0_{00},0}, \ket{2_{20},2},\ket{2_{02},2}$ and the respective Rabi
      frequencies connecting the levels are shown as $\Omega_l, \Omega_u$. The dashed horizontal
      line denotes the instantaneous frequency $\omega(t)$ of the optical centrifuge and appropriate
      detunings are given as $\Delta_l,\Delta_u$.}
	\label{fig:3lvl}
\end{figure}

To show how the optical centrifuge's acceleration rate and field strength guides the rotational
excitation path in near-oblate asymmetric top molecules we build a simple model of three rotational
energy levels: $\ket{J_{k_ak_c}, m} = \ket{0_{00},0}, \ket{2_{20},2},\ket{2_{02},2}$. The
appropriate energy levels together with Rabi frequencies $\Omega_l, \Omega_u$ connecting these
levels are displayed in \autoref{fig:3lvl}.

\begin{figure}
   \includegraphics[width=\linewidth]{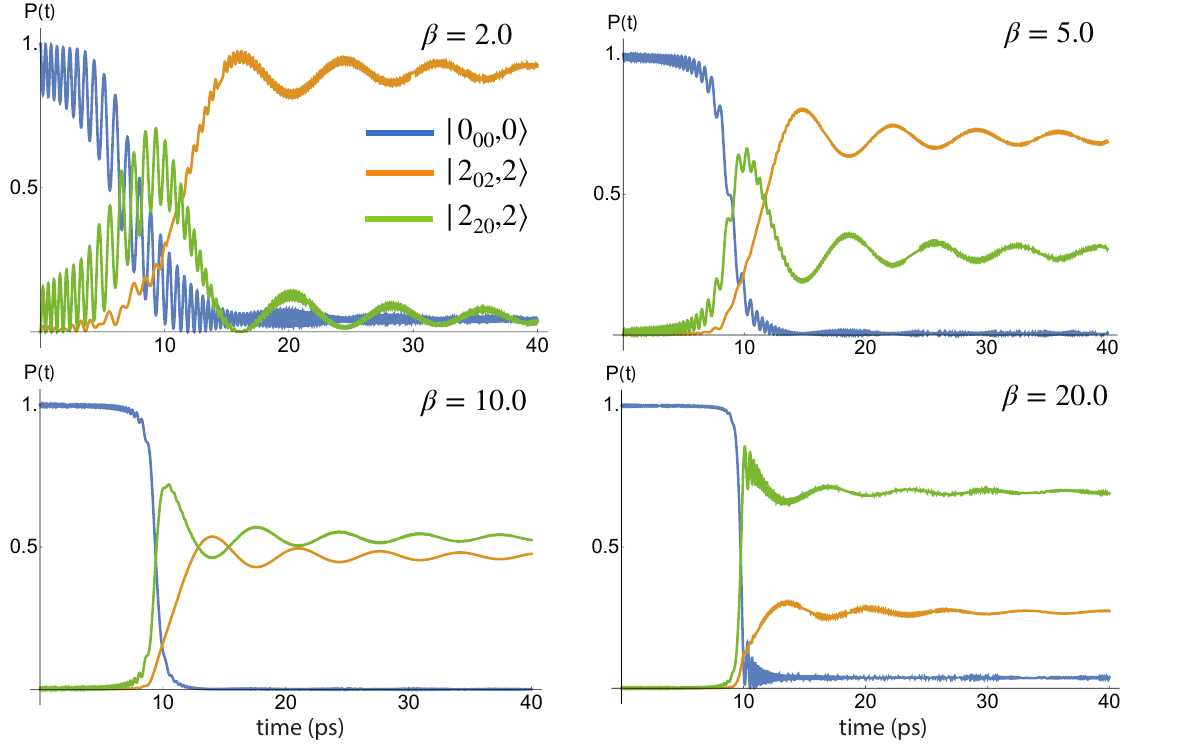}%
   \caption{Calculated population time-profiles for the three-level system defined in
      \autoref{fig:3lvl} at different values of the centrifuge's acceleration rate
      $\beta=2.0,5.0,10.0,20.0$~(arb. unit).}
	\label{fig:3lvl-solutions}
\end{figure}

Numerical solutions to the time-dependent Schrödinger equation for this system are shown in
\autoref{fig:3lvl-solutions} for four different values of the centrifuge's acceleration rate
$\beta$; where $\omega(t)=\beta{}t$.
In near-oblate molecules the transition moment to the upper energy level $\ket{2_{20},2}$ dominates
over the transition moment to the lower energy level $\ket{2_{02},2}$, see~\autoref{fig:transitions}.
The ratio of these transition moments depends on the molecule. Here we adopted the ratio
$\Omega_l/\Omega_u=3.0$ observed in \imidazole. The calculated population time-profiles shown in
\autoref{fig:3lvl-solutions} clearly suggest that at high $\beta$ values the upper energy level is
predominantly populated, whereas low acceleration rates of the centrifuge field prefer the lower
energy level. Intermediate $\beta$ values create a coherent mixture of both excited states.

\section{Cumulative populations}
\label{app-B}
\begin{figure*}
   \includegraphics[width=\linewidth]{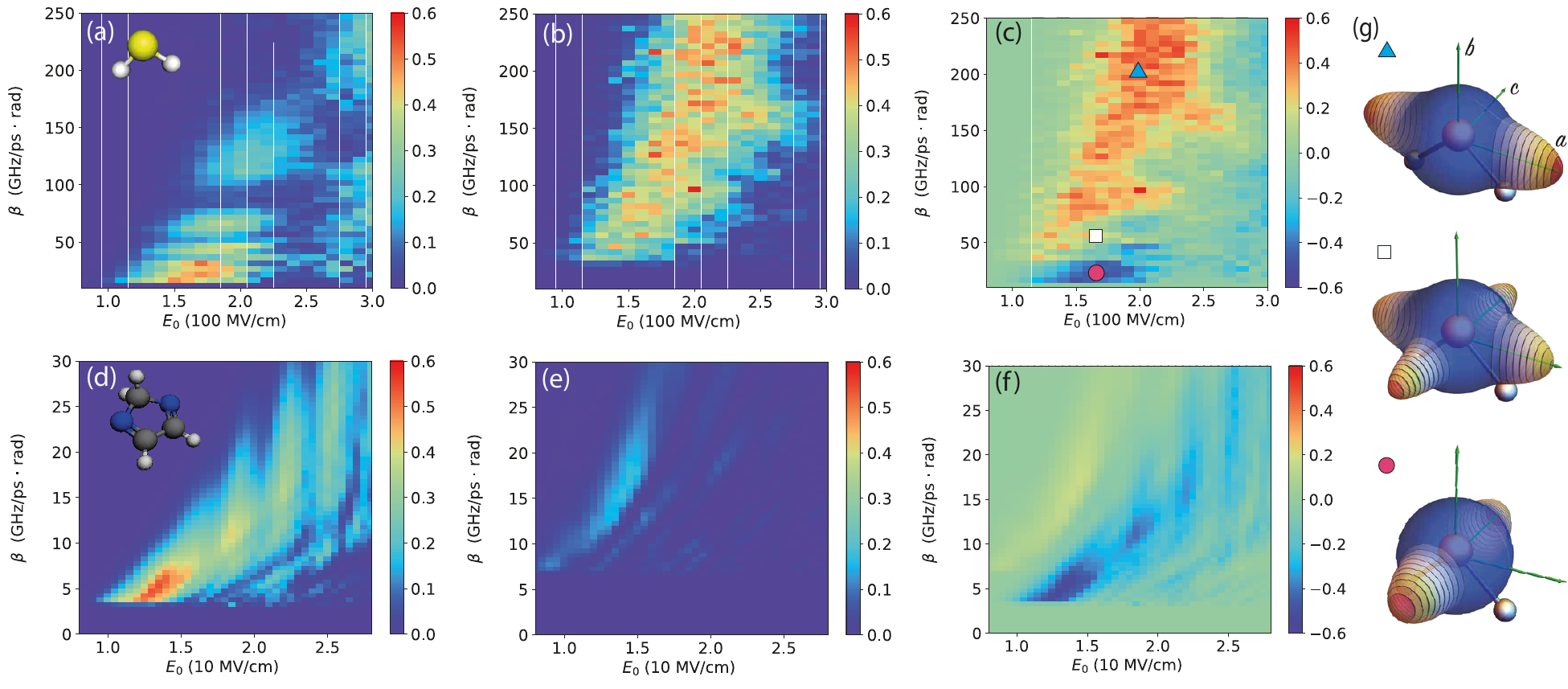}%
   \caption{Cumulative populations of the \ket{c} and \ket{a} principal rotation states as functions
      of the centrifuge's peak field strength $E_0$ and acceleration $\beta$, plotted for \dds in
      (a) and (b), respectively, and for \imidazole in (d) and (e). Corresponding cumulative
      population differences between $\ket{c}$ and $\ket{a}$ states are plotted in (c) and (f). For
      clarity, these populations were only plotted for $J\geq10$ for \dds and for $J\geq20$ for
      2H-imidazole. (g) 3D probability distributions for the molecular-frame rotation axes of \dds
      are showing $a$, $a/c$, and $c$ rotation generated with three different sets of centrifuge
      parameters, marked in panel (c) by a triangle, a square, and a circle, respectively. }
	\label{fig:results}
\end{figure*}
In \autoref{fig:results} we show results of quantum-mechanical calculations of rotational dynamics
of \dds and \imidazole in the optical centrifuge. The calculated cumulative populations of the
$|a\rangle$ and $|c\rangle$ principal rotation states are shown in \autoref[a--f]{fig:results}.
These were calculated for $J\geq10$ for \dds and $J\geq20$ for \imidazole in field-free conditions
after the 150~ps centrifuge pulse was turned off. The populations are plotted as functions of the
centrifuge peak field $E_0$ and the rate of acceleration $\beta$. \autoref[c,~f]{fig:results} show
the population inversion between \ket{a} and \ket{c} states. In both molecules, small acceleration
rates populate mainly the \ket{c} states. With increasing acceleration rate the optical centrifuge
populates more preferably the \ket{a} states. The relation between the wavepacket composition and
the probability of the rotation-axis orientation in the molecular frame is displayed in
\autoref[g]{fig:results} for three selected sets of the centrifuge parameters. The lower overall
excitation efficiency in \imidazole than in \dds is caused by its higher density of states, already at small $J=2,4$ values, which
results in further bifurcations and eventually losses from
the centrifuge.

\section{Molecular rotability}
\label{app-C}
The purpose for the molecular rotability introduced in~\eqref{eq:rotability} is to qualitatively
characterize the relation between the optical centrifuge field parameters $(\beta, E_0)$ and those
of the molecule, the polarizability tensor $\alpha_{ij}$, moments of inertia tensor $I_i$, and
dipole transition moments $\mu_{kl}$ with the composition of the rotational wavepacket.

A linear molecule in the optical centrifuge field follows the classical equation of motion
$I\ddot{\theta} = - \beta I -U_0\sin2\theta$, where
$U_0 = \frac{1}{4}E_0(\alpha_{||}-\alpha_{\perp})$ is the pendular potential depth. Here
$\alpha_{||}$ is the electronic polarizability along the long molecular axis and $\alpha_{\perp}$ is
the perpendicular component of the polarizabiilty. $I$ is the moment of inertia and $\theta$ denotes
the angle between the molecular long axis and the electric field vector of length $E_0$.

In order to capture the likelihood of trapping the molecule in the pendular potential well created
by the centrifuge field one needs to compare the angular acceleration of the centrifuge field with
the angular acceleration of the molecule caused by this field, as shown below:
\begin{equation}
	\beta < \frac{K_{\parallel,max}}{I} = - \frac{\nabla_{\theta} V(\theta)|_{max}}{I} \approx \frac{U_0}{I}
	\label{eq:angular-acceleration}
\end{equation}
where $V(\theta)$ is the pendular potential and $K_{\parallel,max}$ is the maximum torque applied to
the molecule. The trapping condition given in~\eqref{eq:angular-acceleration} can be also directly
inferred from the classical equation of motion:
$\frac{\ddot{\theta}+\beta}{\beta} = -\frac{U_0}{I\beta}\sin2\theta$. Minding the condition in
\eqref{eq:angular-acceleration}, a single parameter which quantifies the \textit{trapability} of a
linear molecule in the centrifuge field can be written as:
\begin{equation}
	S^\text{(1D)} := \frac{E_0^2}{\beta}\frac{\alpha_{\parallel}-\alpha_{\perp}}{I}
	\label{eq:rotability-1D}
\end{equation}
If the acceleration $\beta$ of the field is lower than the acceleration (response) of the molecule
in this field, the molecule is effectively trapped in the pendular potential well and becomes
centrifuged, \ie, rotationally excited. Thus, for $S^\text{(1D)}\gg1$ the rotational excitation
should be efficient.

What distinguishes asymmetric-top molecules from linear molecules is that in the former there is
more than one possible rotational excitation pathway. In linear molecules the angular momentum is
always perpendicular to the molecule's long axis. In asymmetric-top molecules the excitation can
be accompanied by a change in the molecular-frame angular momentum along the $a$, $b$, or $c$
principal axis. Therefore, the 3D-rotability measure must distinguish rotational excitations along
the different principal inertia axes. For this reason a generalization of \eqref{eq:rotability-1D}
to asymmetric-top molecules must capture the details of the full 3D molecule-field-interaction
potential given by
\begin{widetext}
   \begin{equation}
      \begin{aligned}
         V(\theta,\phi,\chi,t) &= \frac{1}{4}\epsilon_0^2\cos^2(\omega
         t)\left[-\frac{2}{3}(\alpha_{xx}+\alpha_{yy}+\alpha_{zz})D_{00}^{(0)*}
         \right. -e^{2i\beta t^2}(\alpha_{xz}-i\alpha_{yz})D_{-2,-1}^{(2)*}
         -e^{-2i\beta t^2}(\alpha_{xz}-i\alpha_{yz})D_{2,-1}^{(2)*}+ \\
         &+e^{2i\beta t^2}(\alpha_{xz}+i\alpha_{yz})D_{-2,1}^{(2)*} +e^{-2i\beta
            t^2}(\alpha_{xz}+i\alpha_{yz})D_{2,1}^{(2)*}
         +\sqrt{\frac{2}{3}}(\alpha_{xz}-i\alpha_{yz})D_{0,-1}^{(2)*}
         -\sqrt{\frac{2}{3}}(\alpha_{xz}+i\alpha_{yz})D_{0,1}^{(2)*}+\\
         &+\frac{1}{\sqrt{6}}e^{2i\beta
            t^2}(\alpha_{xx}+\alpha_{yy}-2\alpha_{zz})D_{-2,0}^{(2)*}
         +\frac{1}{\sqrt{6}}e^{-2i\beta t^2}(\alpha_{xx}+\alpha_{yy}-2\alpha_{zz})D_{2,0}^{(2)*}
         -\frac{1}{3}(\alpha_{xx}+\alpha_{yy}-2\alpha_{zz})D_{0,0}^{(2)*} \\
         & -\frac{1}{2}e^{2i\beta t^2}(\alpha_{xx}-2i\alpha_{xy}-\alpha_{yy})D_{-2,-2}^{(2)*}
         -\frac{1}{2}e^{2i\beta t^2}(\alpha_{xx}+2i\alpha_{xy}-\alpha_{yy})D_{-2,2}^{(2)*}
         +\frac{1}{\sqrt{6}}(\alpha_{xx}-2i\alpha_{xy}-\alpha_{yy})D_{0,-2}^{(2)*} + \\
         &+\frac{1}{\sqrt{6}}(\alpha_{xx}+2i\alpha_{xy}-\alpha_{yy})D_{0,2}^{(2)*}
         -\frac{1}{2}e^{-2i\beta t^2}(\alpha_{xx}-2i\alpha_{xy}-\alpha_{yy})D_{2,-2}^{(2)*}
         \left. -\frac{1}{2}e^{-2i\beta t^2}(\alpha_{xx}+2i\alpha_{xy}-\alpha_{yy})D_{2,2}^{(2)*} \right] \\
      \end{aligned}
      \label{eq:voc}
   \end{equation}
\end{widetext}
where $D_{\Delta{M},\Delta{k}}^{(J)*} $ is the complex-conjugated Wigner $D$ matrix and
$\theta,\phi,\chi$ are the Euler angles. The elements of static polarizability tensor $\alpha_{xx}$,
$\alpha_{xy}$, $\alpha_{xz}$, $\alpha_{yy}$, $\alpha_{yz}$, and $\alpha_{zz}$ refer to the
molecule-fixed principal-axis-of-inertia frame. Accordingly, the 3D molecular rotability must be a
vector with its three components denoting the net affinity of the molecule to rotate about the
three-respective principal axes as a function of the centrifuge field parameters. In order to
heuristically derive the components of the 3D molecular rotability we follow the condition given
in~\eqref{eq:angular-acceleration} and calculate the torques created along the respective principal
inertia axes.

Here we note that rather than deriving a condition for the molecule to stay in the centrifuge
potential energy trap, we aim at providing a measure, which weighs the relative affinity of
the principal polarisability axes to the polarisation plane of the centrifuge, i.e. which of the principal rotation axes is the most
likely. Without referring to complex classical dynamics of the rigid 3D
molecule in rotating electric field we give below a simplified justification for molecular rotability.

First note that for each of three possible axes of rotation there is an effective torque caused
by the interaction with the cenitrfuge field. The relative magnitude of the torque
$K^{(\lambda)}_{||}$ created along the $\lambda$-axis (aligned with $Z$-axis) to the torque
$K^{(\lambda)}_{\perp}$ along an axis perpendicular to $Z$ (in the $XY$ plane) informs about the
affinity of the molecule to stay aligned along the given rotation axis. We imagine three
arrangements, in which the molecule rotates about the $a$-, $b$- or $c$- axis aligned along the
laboratory $Z$-axis and calculate the ratio of the aforementioned torques:
\begin{equation}
   \begin{aligned}
      \xi_{\lambda} = \frac{K^{(\lambda)}_{||}}{K^{(\lambda)}_{\perp}}, \qquad \lambda = a,b,c
   \end{aligned}
   \label{eq:torques}
\end{equation}
with
\begin{equation}
   \begin{aligned}
      K^{(\lambda)}_{||} =\left|\frac{\partial V(t)}{\partial \phi}\right|\\
      K^{(\lambda)}_{\perp} =\left| \frac{\partial V(t)}{\partial \theta}\right| \\
   \end{aligned}
   \label{eq:torque-def}
\end{equation}
where $V_i(t)$ denotes the potential given in \eqref{eq:voc} and $\theta, \phi$ are the azimuthal
and polar Euler angles, respectively. With \eqref{eq:voc} the parallel and perpendicular torque
components are given as
\begin{equation}
   \begin{aligned}
      K^{(\lambda)}_{||} =\frac{\epsilon_0^2}{8}\left|(\alpha_{xx}-\alpha_{yy}) d_{-2,2}^{(2)}(\theta)\right|\\
      K^{(\lambda)}_{\perp} =\frac{\epsilon_0^2}{4\sqrt{6}}\left|\left((\alpha_{xx}-\alpha_{zz})+(\alpha_{yy}-\alpha_{zz})\right)
         \frac{\partial
            d_{-2,0}^{(2)}(\theta)}{\partial \theta}\right| \\
   \end{aligned}
   \label{eq:torque}
\end{equation}
where $d_{\Delta{m},\Delta{k}}^{(J)}$ are elements of the real-valued Wigner \textit{small-d}
matrix. We dropped any contributions from the off-diagonal elements of the electronic polarizability
tensor. In majority of small molecules the principal inertia axes frame and the frame in which the
electronic polarizability is diagonal nearly overlap, which means that the off-diagonal elements of
the electronic polarizability in the principal inertia axes frame are very small compared to the
diagonal elements. We note that $d_{-2,0}^{(2)}(\theta) = \sqrt{\frac{3}{8}}\sin^2\theta$ and
$d_{-2,-2}^{(2)}(\theta) =\cos^4\theta/2$ and are bound from above by $1$ and $\sqrt{\frac{3}{8}}$,
respectively. Therefore the ratio of maximum torques is approximately given as
\begin{equation}
   \begin{aligned}
      \xi_{\lambda} \approx \frac{ \left|\alpha_{xx}-\alpha_{yy}\right|}{ \left|(\alpha_{xx}-\alpha_{zz})+(\alpha_{yy}-\alpha_{zz})\right|}
   \end{aligned}
   \label{eq:maxratio}
\end{equation}
where $\lambda = a,b,c$. Based on \eqref{eq:maxratio} we propose a definition for the generalized
trapability (trapping ability), which accounts for these maximum torques, in the following form:
\begin{equation}
	R_i=\frac{C_i\left|\alpha_{jj}-\alpha_{kk}\right|}{C_j\left|\alpha_{kk}-\alpha_{ii}\right|+C_k\left|\alpha_{jj}-\alpha_{ii}\right|}, \quad
	i=a,b,c
	\label{eq:trapability}
\end{equation}
This definition comprises the relative pendular potential depths projected along different principal
inertia axes. Here, $\alpha_{ii}$ are the diagonal components of the electronic polarizability and
$C_i=A,B,C$ for $i=a,b,c$ are the rotational constants. In simple terms the, 3D trapability is
constructed from 1D trapabilities and reflects the molecule's affinity to rotate about a given
inertial axis as compared to the other two axes. We assume that the principal axes of inertia system
is parallel to the electronic polarizability axis system, which is a very good approximation for
\dds and \imidazole.

Another factor that contributes to the final populations of rotational states upon rotational
excitation with the optical centrifuge are the transition-dipole moments to the upper and the lower
rotational excitation branches. The interaction of molecules with non-resonant chirped fields such
as the optical centrifuge can be qualitatively described with the Landau-Zener
model~\cite{Vitanov:PRA53:4288}. The branching ratio for the rotational excitation along the upper
or lower excitation path, see~\autoref{fig:energy-levels}, can be estimated as the ratio of the
Landau-Zener populations:
\begin{equation}
	Q_{ij}=P_i/P_j
	\label{eq:transferability}
\end{equation}
where $P_i=1-\exp\left(-\pi\Omega_i^2/(4\beta)\right)$~\cite{Vitanov:PRA53:4288} and the Rabi
frequency $\Omega_i=\mu_{i}E_0^2$, which depends on the transition moment $\mu_{i}$ from the rotational ground
state to one of the \ket{i} ($i=a,b,c$) principal rotation states and $E_0$ is the electric field
strength. The quantum-mechanical population \emph{transferability} $Q_{ij}$ relates the centrifuge
field parameters to the anticipated branching ratio for the population transfer to respective
rotational states.

Putting together the molecular trapability and transferability we can assume that the molecular
trapability given in \eqref{eq:trapability} serves as a scaling factor for the rotational excitation
probability to the upper-energy-levels branch ($\ket{a}$ states) versus the lower-energy-levels
branch ($\ket{c}$ states), quantified by the trasferability $Q_{ij}$.

Finally the 3D molecular rotability can be written as
\begin{equation}
	S_i \coloneqq R_i (Q_{ji}+Q_{ki})  \qquad i,j,k=a,b,c
	\label{eq:rotability-app}
\end{equation}
In the prolate ($A>B=C$) and oblate ($A=B>C$) symmetric-top limits the rotability vector becomes
\begin{align}
	\vec{S}_{prolate} &\rightarrow \begin{pmatrix} 0 \\ Q \\ Q \end{pmatrix} \\
	\vec{S}_{oblate} &\rightarrow \begin{pmatrix} Q \\ Q \\ 0 \end{pmatrix}
\end{align}
where $Q = Q_{ba}+Q_{ca}$. Thus, in prolate (oblate) symmetric tops there is no possibility for
creating torque in the $bc$ ($ab$) molecular plane, \ie, the probability for rotational excitation
about the $a$ ($c$) axis through the electronic polarizability is $0$.

\begin{figure}
   \includegraphics[width=\linewidth]{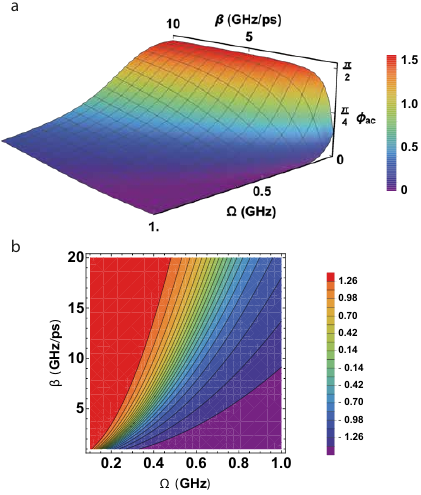}%
   \caption{(a) Graphical representation of the relation between the molecular rotability components
      $S_a$ and $S_c$ in the form of the angle: $\phi_{ac}=\arctan(\frac{S_a}{S_c})$. $\beta$ is the
      centrifuge acceleration rate and $\Omega$ is the Rabi frequency for a unit transition dipole
      moment. (b) Molecular rotability map with predicted population inversion between the $a$ and
      $c$ rotational-excitation branch. The plotted function is
      $\arctan(\frac{S_a}{S_c}-\frac{S_c}{S_a})$.}
   \label{fig:rotability}
\end{figure}
In \autoref{fig:rotability} we show an example of the molecular rotability components $S_a-S_c$ for
the molecular parameters of \imidazole as a function of the centrifuge electric field strength $E_0$
and acceleration rate $\beta$. We see that at low acceleration rates the $S_c$ component dominates
over the $S_a$ component, which suggests that the molecule will most likely follow the $c$ axis
rotational excitation branch. The rotability measure covers all wavepackets, thus the rotational
wavepacket can be expressed in terms of the rotability vector components:
\begin{equation}
	\ket{\psi} = \frac{2}{\pi}\left(\arctan^2\left(\frac{S_a}{S_c}\right) \ket{a}
	+ \sqrt{1-\arctan^2\left(\frac{S_a}{S_c}\right)}\ket{c}\right)
	\label{eq:wavepacket}
\end{equation}
Geometrically, \eqref{eq:wavepacket} represents a map between the combined molecular parameters and
the parameters of the centrifuge onto a line connecting $\ket{a}$ and $\ket{c}$ states. Rotational
states can be classified according to the respective average values of the molecule-fixed angular
momentum, and can be represented on a triangle, as shown in the inset of~\autoref{fig:energy-levels} .
Vertices of the triangle denote essentially pure $\ket{a}$, $\ket{b}$, $\ket{c}$ states. In principle, the
composition of the final rotational wavepacket created by the interaction with external fields can
be mapped onto this triangle. However, because the rotational excitation about the intermediate $b$
axis is very improbable, due to low values of appropriate transition moments, the molecular
rotability practically maps the space of centrifuge's parameters $E_0$ and $\beta$ onto the edge of
the rotational states triangle connecting the $\ket{a}$ and $\ket{c}$ vertices. For this reason the
image of the molecular rotability map for the optical centrifuge is approximately one-dimensional.

\autoref[b]{fig:rotability} reflects the magnitude of population inversion in the rotational $a$ or
$c$ branches. It shows a qualitative agreement with the full quantum-mechanical optical centrifuge
calculations for \imidazole, shown in~\autoref{fig:populations}.

\section{Analysis of rotational wavepackets for \dds: probability densities and alignment cosines}
\label{app-D}
\begin{figure}
	\includegraphics[width=\columnwidth]{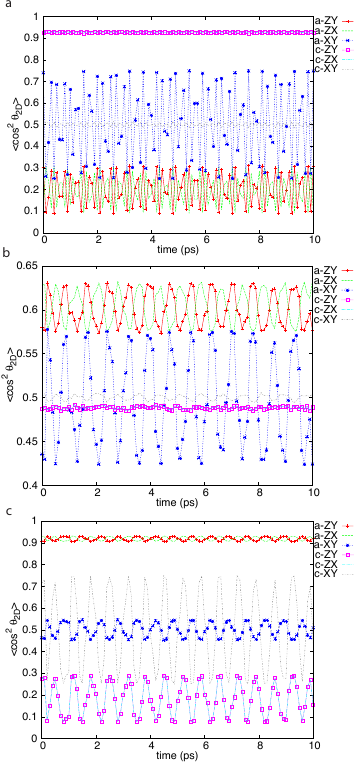}%
	\caption{Calculated 2D alignment-cosine time-profiles for wavepackets generated with an optical
		centrifuge with field strength $E_0=2\cdot10^{8}$~V/cm and three different acceleration rates
		(a--c) \betarate{30,115,230}, respectively. The centrifuge-field release times were 228~ps,
		133~ps, 66~ps, respectively, see~\autoref{fig:PDF}. The alignment cosines measure
		the degree of alignment of the principal axes of inertia of \dds with respect to the
		laboratory axes defined by the centrifuge field. For instance $a-ZY$ is the 2D alignment
		cosine calculated with respect to the laboratory $Z$ axis of the projection of the molecular
		$a$ axis onto the $ZY$ detector plane.}
	\label{fig:cosines}
\end{figure}
The approximate wavepackets generated with an optical centrifuge at three different acceleration
rates \betarate{30,~115,~230} and field strength \mbox{$E_0=2\cdot10^{8}$~V/cm}, shown in \autoref{fig:PDF}, are given as
\begin{align}
	\label{eq:wf-30}
	\ket{\psi}_{\beta=30}  \sim &\; 0.55e^{-i\omega_{12a} t}\ket{12,a} +  0.55 e^{-i\omega_{14a} t}\ket{14,a}
	\\
	\label{eq:wf-115}
	\ket{\psi}_{\beta=115} \sim &\; 0.32 e^{-i\omega_{28c} t}|28,c\rangle +0.32e^{-i\omega_{30c} t} |30,c\rangle \notag  \\
	& +0.32e^{-i\omega_{2c} t} |2,c\rangle + 0.61 e^{-i\omega_{14a} t}|14,a\rangle
	\\
	\label{eq:wf-230}
	\ket{\psi}_{\beta=230} \sim &\; 0.36 e^{-i\omega_{12c} t} |12,c\rangle +  0.36 e^{-i\omega_{14c} t}|14,c\rangle
\end{align}
where $\ket{J,a/c}$ are the asymmetric top wavefunctions with total angular momentum $J$ and the
molecular-frame angular momentum nearly aligned with $a$ or $c$ axis, respectively. All states
have $M=J$, \ie, the laboratory-frame angular momentum is aligned along the $Z$ axis.

Time-evolution of the rotation-axis 3D probability ($1-e^{-1}$ cutoff) in the above wavepackets is
displayed in the animated movie files attached: \texttt{D2S-$\beta$-a-axis.mov}, \texttt{D2S-$\beta$-b-axis.mov},
\texttt{D2S-$\beta$-c-axis.mov} for the $a$,
$b$ and $c$- principal inertia axis, respectively. Joint 3D probability plots for the deuterium
nuclei are given in files \texttt{D2S-$\beta$-D-atoms.mov}. In these movie files a classical-like rotation of
the probability is visible for all three wavepackets given in \eqref{eq:wf-30}-\eqref{eq:wf-230}.

These rotational wavepackets can be straightforwardly detected with the use of velocity-map imaging
(VMI) of fragments produced through multiple ionization by ultrashort laser pulses followed by
Coulomb explosion~\cite{Stapelfeldt:PRA58:426,Corkum:ARPC48:387}. Files named
\texttt{D2S-$\beta$-vmixz.mov}, \texttt{D2S-$\beta$-vmiyz.mov} and \texttt{D2S-$\beta$-vmixy.mov}
present time-evolution of the VMI images simulated for the deuterium ion fragments, assuming axial
recoil. The detector is placed in the laboratory-fixed $XZ$, $YZ$ and $XY$ plane, respectively,
where $Z$ is the laser propagation direction.
Only the first two setups ($XZ$, $YZ$) are routinely
implemented in the experiment. The VMI images can be quantitatively characterized by calculating the
degree of alignment of selected atoms or principal axes of inertia with respect to the laboratory
$X,Y,Z$ axes. The alignment cosine values, which reconstruct the experimental VMI images, can be
calculated from the positions of selected atoms or molecular axes $x,y,z$ in the laboratory-frame
$X,Y,Z$ using
\begin{equation}
	\cost = \bra{\psi}  \frac{Z^2}{X^2 + Z^2}  \ket{\psi}
	\label{eq:cos2d}
\end{equation}
where $\theta_{2D}$ is an angle in the detector plane $XZ$ between the detector's $Z$ axis and the
D-S bond in \dds at the time of the Coulomb explosion. In this work the alignment cosine values were
calculated with Monte Carlo sampling, using $10^6$ sampling points, of the rotational wavefunction
$\ket{\psi}$ at a given time. \autoref{fig:cosines} displays values of $\cost$
calculated for the vectors pointing along the molecular $a$ and $c$ axis, respectively, relative to
the laboratory-fixed planes: $XY$, $XZ$ and $YZ$.

\autoref{fig:cosines} shows a high ($\larger0.9$) degree of permanent alignment of the $c$ axis
along the laboratory fixed $Z$ axis for the wavepacket given in \eqref{eq:wf-30} generated with
\betarate{30} and field strength $E_0=2\cdot10^{8}$~V/cm, see~\autoref{fig:PDF}. The
classical-like rotational motion of the $ab$ molecular plane about the $c$ axis is reflected in the
oscillations of the blue line ($a$--$XY$) in \autoref[a]{fig:cosines}. Small amplitude oscillations
of the red and green $a$--$ZY$ and $a$--$ZX$ lines near $\cost=0.2$ suggest slight nutation of the
$a$ axis out of the $XY$-plane in the laboratory frame. Wavepackets shown in
\eqref{eq:wf-30}--\eqref{eq:wf-230} exhibit persistent 3D alignment of the molecule in the frame
rotating with the frequency of the coherent classical-like rotation of the whole wavepacket. Period
of this uniform rotation is in the order of few picoseconds, which is long enough for most of
imaging ultrafast dynamics experiments.

\bibliography{string,cmi}
\end{document}